\documentstyle[11pt,cite]{article}
\newcommand{\non}{\nonumber \\}
\newcommand{\be}{\begin{equation}}
\newcommand{\ee}{\end{equation}}
\newcommand{\bea}{\begin{eqnarray}}
\newcommand{\eea}{\end{eqnarray}}
\newcommand{\lp}{\left (}
\newcommand{\rp}{\right )}
\newcommand{\lb}{\left \{}

\newcommand{\lbr}{\left [}
\newcommand{\rbr}{\right ]}
\newcommand{\ld}{\left .}
\newcommand{\rd}{\right .}
\newcommand{\freme}{\frac{4!}{\bar{\cal M}_4}}
\newcommand{\freae}{\frac{4!}{a_4^{(n)}}}
\newcommand{\ve}[1]{{\bf #1}}
\newcommand{\rhok}{\rho_{\ve{k}}}
\newcommand{\rhomk}{\rho_{-\ve{k}}}
\newcommand{\omk}{\omega_{\ve{k}}}

\newcommand{\cM}{{\cal M}}
\newcommand{\bcMe}{\bar{\cal M}_4}
\newcommand{\cW}{{\cal W}}
\newcommand{\sli}{\sum\limits}
\newcommand{\ili}{\int\limits}
\newcommand{\btPhik}{\beta\tilde{\Phi}(k)}
\newcommand{\half}{\frac{1}{2}}

\newcommand{\sect}[1]{\section{#1}\setcounter{equation}{0}}
\begin{document}


\begin{center}
{\bf 3D ISING SYSTEM IN AN EXTERNAL FIELD. RECURRENCE RELATIONS FOR THE 
ASYMMETRIC $\rho^6$ MODEL}
\end{center}

\begin{center}       
{\sc I.V.Pylyuk, M.P.Kozlovskii}
\end{center} 

\begin{center}
{\it Institute for Condensed Matter Physics  \\
of the National Academy of Sciences of Ukraine, \\
1~Svientsitskii Str., UA-79011 Lviv, Ukraine} \\
E-mail: piv@icmp.lviv.ua
\end{center}

\vspace{0.5cm}

{\small
The 3D one-component spin system in an external magnetic field is studied 
using the collective variables method. The integration of the partition 
function of the system over the phase space layers is performed in the 
approximation of the sextic measure density including the even and the odd 
powers of the variable (the asymmetric $\rho^6$ model). The general 
recurrence relations between the coefficients of the effective measure 
densities are obtained. The new functions appearing in these recurrence 
relations are given in the form of a convergent series.
}

\vspace{0.5cm}

\sect{Introduction}

The object of investigation is the Ising model on a simple cubic lattice 
with an exponentially decreasing interaction potential. The Ising model, 
which is simple and convenient for mathematical analysis, is widely used in 
the theory of phase transitions for an analysis of properties of various 
magnetic and nonmagnetic systems (binary mixtures, lattice model of liquids, 
ferromagnets, etc.). 

The three-dimensional (3D) Ising system in an external magnetic field will 
be studied here based on the collective variables (CV) method 
\cite{ymo287}. This original method can be extended to the liquid-gas system 
\cite{rev9295}, multicomponent fluids \cite{rev9395}. In this case, the 
method makes it possible to use a convergent series by applying the solution 
of the 3D Ising model in contrast to the asymptotic $\epsilon$-expansions 
(see, for example, \cite{vs180,n281}). 

The CV method allows one to calculate the partition function of the  
Ising model and to obtain not only the universal quantities but also 
analytic expressions for nonuniversal characteristics as functions of the 
microscopic parameters of the system. The statistical description of the 
spin system behaviour is performed in real 3D space on the microscopic 
level. The main results obtained in the CV method for the 3D systems in the 
absence of an external field are presented in \cite{ymo287,p398e}.  

An important condition for describing the system properties by the CV 
method is the use of non-Gaussian measure densities \cite{ymo287}. Such a 
measure density at a zero external field is represented as an exponential 
function of the CV, the argument of which contains, along with the quadratic 
term, higher even powers of the variable with the corresponding coupling 
constants. The confinement to the $6$th power of the CV in the expression 
for a measure density corresponds to the $\rho^6$ model. This model provides 
a more adequate quantitative description of the 3D Ising system behaviour 
in comparison with the $\rho^4$ model (see \cite{kpd297,p599e,p699e}).

In the case of a nonzero external field, the non-Gaussian measure density 
includes the terms proportional to odd powers of the CV in addition to the 
terms proportional to even powers. The partition function of the 3D 
one-component system in an external field is presented below in the 
approximation of the sextic measure density, which involves the even and 
odd powers up to the $6$th power of the variable (the asymmetric $\rho^6$ 
model). A partition function functional is similar to the Ginzburg-Landau 
functional (see, for example, \cite{mmo176e}). We calculate the partition 
function by successively integrating its expression over the short-wave or 
the so-called unimportant variables. The corresponding renormalization 
group (RG) transformation can be 
related to the Wilson type \cite{rev14674}. The 
general recurrence relations (RR) corresponding to the asymmetric $\rho^6$ 
model and the new special functions entering the RR are considered.

In our earlier works (see, for example, \cite{kpd297,p599e,p699e}), an 
infinitely weak external field was introduced in the course of calculation 
of the contribution from the long-wave modes of the spin moment density 
oscillations to the 3D Ising system thermodynamic characteristics. In this 
paper, we introduce an external field in the Hamiltonian from the outset. 
Such an approach leads to the appearance of odd powers of the CV in the 
expression for the partition function and makes it possible to describe a 
lot of quantities (cumulants, initial coefficients of the partition 
function, etc.) as functions of an external field.

\sect{Recurrence relations for the asymmetric $\rho^6$ mo\-del}

We consider the Ising system on a simple cubic lattice with period 
$c$. The Hamiltonian of the system has the form
\be
H=-\half~\sli_{\ve{i},\ve{j}}\Phi(r_{\ve{i}\ve{j}})\sigma_{\ve{i}} 
\sigma_{\ve{j}}-h\sli_{\ve{i}}\sigma_{\ve{i}}.
\label{smt2d1}
\ee
Here $h$ is an external field, $r_{\ve{i}\ve{j}}$ is the distance between
particles at the sites $\ve{i}$ and $\ve{j}$, $\sigma_{\ve{i}}$ is the 
operator of the $z$-component of spin at the $\ve{i}$th site, having two 
eigenvalues +1 and $-1$. The interaction potential is an exponentially 
decreasing function
\be
\Phi(r_{\ve{i}\ve{j}})=A\exp\lp-\frac{r_{\ve{i}\ve{j}}}{b}\rp,
\label{smt2d2}
\ee
where $A$ is a  constant, $b$ is the radius of effective interaction.
For the Fourier transform of the interaction potential, we use the 
following approximation:
\be
\tilde{\Phi}(k)=\lb
\begin{array}{ll}
\tilde{\Phi}(0)(1-2b^2k^2)  & \mbox{at}~~k\leq B', \\
0 & \mbox{at}~~B'<k\leq B.
\end{array}
\rd
\label{smt2d3}
\ee
Here $B=\pi/c$ is the boundary of the Brillouin half-zone, 
$B'=(b\sqrt 2)^{-1}$ is determined from the condition for the application 
of the parabolic approximation $\tilde{\Phi}(0)(1-2b^2k^2)=0$,
$\tilde{\Phi}(0)=8\pi A(b/c)^3$. At $\tilde{\Phi}(0)=2dJ, 
~b=b_{I}=c/(2\sqrt d)$ ($J$ is the constant of the interaction between the 
nearest neighbours, $d=3$ is the space dimension) for small values of the 
wave vectors $\ve{k}$, the parabolic approximation of the Fourier transform 
of the exponentially decreasing interaction potential corresponds to the 
analogous approximation of the Fourier transform for the interaction 
potential between the nearest neighbours \cite{kpu196}. 

In the CV representation for the partition function of the 3D Ising model
in an external field, we have
\be
Z=\int\exp\lbr\half~\sli_{\ve{k}}\btPhik\rhok\rhomk\rbr J(\rho)~(d\rho)^N,
\label{smt2d4}
\ee
where the summation over the wave vectors $\ve{k}$ is carried out within the 
first Brillouin zone, $\beta=1/(k_BT)$ is the inverse thermodynamic 
temperature, $k_B$ is the Boltzmann constant. The CV $\rhok$ are introduced 
using the relations of the type of an analytic functional for 
operators of spin density oscillation modes 
$\hat\rhok=(\sqrt{N})^{-1}\sum_{\ve{l}}  
\sigma_{\ve{l}}\exp(-i\ve{k}\ve{l})$ \cite{ymo287}. The expression for 
$J(\rho)$ can be written as
\bea
J(\rho)&=&(2\cosh h')^N\int\exp\lbr 2\pi 
i~\sli_{\ve{k}}\omk\rhok+~\sli_{n\geq 1}\lp -2\pi i\rp^{n} 
N^{-(n-2)/2}\times\rd \non
&&\ld\times\frac{\cM_n}{n!}~\sli_{\ve{k}_1,\ldots,\ve{k}_n} 
\omega_{\ve{k}_1}\cdots\omega_{\ve{k}_n}\delta_{\ve{k}_1+\cdots+
\ve{k}_n}\rbr~(d\omega)^N.
\label{smt2d5}  
\eea
Here $\delta_{\ve{k}_1+\cdots+\ve{k}_n}$ is the Kronecker symbol, the 
variables $\omk$ are conjugate to $\rhok$. The cumulants $\cM_n$ are 
functions of a generalized field $h'=\beta h$:
\bea
& & \cM_1=\tanh h'=x, \non
& & \cM_2=1-\tanh^2 h'=y, \non
& & \cM_3=-2xy, \non 
& & \cM_4=-2y^2+4x^2y, \non 
& & \cM_5=16xy^2-8x^3y, \non 
& & \cM_6=16y^3-88x^2y^2+16x^4y.  
\label{smt2d5a}
\eea
It should be noted that (\ref{smt2d5}) is the Jacobian of the transition 
from the set of the $N$ spin variables $\sigma_{\ve{l}}$ to the set of the 
CV $\rhok$ at $h=0$. 

We shall proceed from the expression for the partition function in the 
approximation of the asymmetric $\rho^6$ model. Putting $n=6$ in 
(\ref{smt2d5}) and carrying out integration in (\ref{smt2d4}) with respect 
to the variables $\rhok$ and $\omk$ with indices $B'<k\leq B$, followed by 
the integration with respect to the $N'$ variables $\omk$, we obtain
\bea
Z&=&(2\cosh h')^N2^{(N'-1)/2}e^{a'_0N'}\int\exp\Biggl[ 
-(N')^{1/2}a'_1\rho_0-\half~\sli_{k\leq B'}d'(k)\times \non
&&\times\rhok\rhomk-~\sli_{l=3}^6\frac{a'_l}{l!(N')^{l/2-1}}~
\sli_{k_1,\ldots,k_l
\leq B'}\rho_{\ve{k}_1}\cdots\rho_{\ve{k}_l}\times \non
&&\times\delta_{\ve{k}_1+\cdots+\ve{k}_l}\Biggr]~(d\rho)^{N'},
\label{smt2d6}
\eea
where $N'=Ns_0^{-d},~s_0=B/B'=\pi\sqrt 2 b/c$, and
\be
d'(k)=a'_2-\btPhik.
\label{smt2d7}
\ee
The coefficients 
\bea
& & a'_0=\ln Q(\cM), \qquad Q(\cM)=\pi^{-1}\lp\freme 
s_0^d\rp^{1/4}I_{0c}(\eta'), \non
& & a'_1=-\lp\freme s_0^d\rp^{1/4}C_{1sc}(\eta'), \non
& & a'_k=\lp\freme s_0^d\rp^{k/4}C_{ksc}(\eta'), \qquad k=2,3,4,5,6
\label{smt2d8}
\eea
are functions of a field $h'$ and a ratio $b/c$. In this expressions, 
$\bcMe=-\cM_4$. The set $\eta'=\{\eta'_1, \eta'_2, \eta'_3, \eta'_4, 
\eta'_5, \eta'_6\}$ is the set of initial arguments
\bea
& & \eta'_1=\cM_1\lp\freme\rp^{1/4}s_0^{3d/4}, \qquad  
\eta'_2=\frac{\cM_2}{2}\lp\freme\rp^{1/2}s_0^{d/2}, \non
& & \eta'_3=\frac{\cM_3}{3!}\lp\freme\rp^{3/4}s_0^{d/4}, \qquad  
\eta'_4\equiv 1, \non
& & \eta'_5=\frac{\cM_5}{5!}\lp\freme\rp^{5/4}s_0^{-d/4}, \qquad 
\eta'_6=\frac{\cM_6}{6!}\lp\freme\rp^{3/2}s_0^{-d/2}. 
\label{smt2d9}
\eea
The ~special ~functions ~$C_{lsc}(\eta')$ ~in ~(\ref{smt2d8}) ~are 
~combinations ~of ~the ~functions 
$F_{(2l+1)sc}(\eta')=I_{(2l+1)s}(\eta')/I_{0c}(\eta')$ and 
$F_{2lcc}(\eta')=I_{2lc}(\eta')/I_{0c}(\eta')$:
\bea
C_{1sc}(\eta')&=&F_{1sc}(\eta'), \non 
C_{2sc}(\eta')&=&F_{2cc}(\eta')+F_{1sc}^2(\eta'), \non  
C_{3sc}(\eta')&=&F_{3sc}(\eta')-3F_{2cc}(\eta')F_{1sc}(\eta')- 
2F_{1sc}^3(\eta'), \non
C_{4sc}(\eta')&=&-F_{4cc}(\eta')-4F_{3sc}(\eta')F_{1sc}(\eta')+ 
3F_{2cc}^2(\eta')+ \non
&&+12F_{2cc}(\eta')F_{1sc}^2(\eta')+6F_{1sc}^4(\eta'), \non
C_{5sc}(\eta')&=&-F_{5sc}(\eta')+5F_{4cc}(\eta')F_{1sc}(\eta')+ 
10F_{3sc}(\eta')F_{2cc}(\eta')+ \non 
&&+20F_{3sc}(\eta')F_{1sc}^2(\eta')-30F_{2cc}^2(\eta')F_{1sc}(\eta')- \non 
&&-60F_{2cc}(\eta')F_{1sc}^3(\eta')-24F_{1sc}^5(\eta'), \non
C_{6sc}(\eta')&=&F_{6cc}(\eta')+6F_{5sc}(\eta')F_{1sc}(\eta')-
15F_{4cc}(\eta')F_{2cc}(\eta')- \non
&&-30F_{4cc}(\eta')F_{1sc}^2(\eta')+10F_{3sc}^2(\eta')- \non
&&-120F_{3sc}(\eta')F_{2cc}(\eta')F_{1sc}(\eta')-
120F_{3sc}(\eta')F_{1sc}^3(\eta')+ \non
&&+30F_{2cc}^3(\eta')+270F_{2cc}^2(\eta')F_{1sc}^2(\eta')+
360F_{2cc}(\eta')F_{1sc}^4(\eta')+ \non
&&+120F_{1sc}^6(\eta').
\label{smt2d10}
\eea
Here 
\bea
& & I_{(2l+1)s}(\eta')=\ili_0^{\infty}t^{2l+1}f_s(t)~dt, \qquad  
I_{2lc}(\eta')=\ili_0^{\infty}t^{2l}f_c(t)~dt, \non 
& & l=0,1,2,3, 
\label{smt2d11}
\eea
and
\bea
& & f_s(t)=\sin(\eta'_1t-\eta'_3t^3+\eta'_5t^5)\exp(-\eta'_2t^2-
\eta'_4t^4-\eta'_6t^6), \non  
& & f_c(t)=\cos(\eta'_1t-\eta'_3t^3+\eta'_5t^5)\exp(-\eta'_2t^2-
\eta'_4t^4-\eta'_6t^6).  
\label{smt2d12}
\eea

We use the method of layer-by-layer integration of (\ref{smt2d6}) 
with respect to the variables $\rhok$ proposed by I.R. Yukhnovskii 
\cite{ymo287}. The integration begins from the variables $\rhok$ with  
large values of $k$ (of the order of the Brillouin half-zone boundary). The 
phase space of the CV $\rhok$ is divided into layers with the division 
parameter $s$. In each $n$th layer (corresponding to the region of wave 
vectors $B_{n+1}<k\leq B_n$, $B_{n+1}=B_n/s$, $s>1$), the Fourier transform 
of the potential $\tilde\Phi(k)$ is replaced by its average value 
$\tilde{\Phi}(B_{n+1},B_n)$. As a result of step-by-step calculation of the 
partition function, the number of integration variables in the expression 
for this quantity decreases gradually. The partition function is then 
represented as a product of the partial partition functions of individual 
layers and the integral of the "smoothed" effective measure density. After 
the integration over the $n+1$ layers of the CV space, we obtain
\bea
Z&=&(2\cosh h')^N2^{(N_{n+1}-1)/2}Q_0Q_1\cdots Q_n[Q(P_n)]^{N_{n+1}}\times 
\non
&&\times\int\cW_6^{(n+1)}(\rho)~(d\rho)^{N_{n+1}}, 
\label{smt2d13} 
\eea
where $N_{n+1}=N's^{-d(n+1)}$, and
\bea
& & Q_0=\lbr e^{a'_0}Q(d)\rbr^{N'}, \quad Q_1=\lbr Q(P)Q(d_1)\rbr^{N_1},
\quad ~\ldots~, \non
& & Q_n=\lbr Q(P_{n-1})Q(d_n)\rbr^{N_n}, \non
& & Q(d_n)=\lp 4!/a_4^{(n)}\rp^{1/4}I_0(h^{(n)}), \non
& & Q(P_n)=\pi^{-1}\lp s^d a_4^{(n)}/C_4(h^{(n)})\rp^{1/4}
I_{0c}(\eta^{(n)}).
\label{smt2d14} 
\eea
The components 
\bea
& & h_1^{(n)}=a_1^{(n)}\lp\freae\rp^{1/4}, \qquad  
h_2^{(n)}=\frac{d_n(B_{n+1},B_n)}{2}\lp\freae\rp^{1/2}, \non
& & h_3^{(n)}=\frac{a_3^{(n)}}{3!}\lp\freae\rp^{3/4}, \qquad  
h_4^{(n)}\equiv 1, \non
& & h_5^{(n)}=\frac{a_5^{(n)}}{5!}\lp\freae\rp^{5/4}, \qquad 
h_6^{(n)}=\frac{a_6^{(n)}}{6!}\lp\freae\rp^{3/2} 
\label{smt2d15}
\eea
of the set of basic arguments in the $n$th layer \\ 
$h^{(n)}=\{h_1^{(n)},
h_2^{(n)}, h_3^{(n)}, h_4^{(n)}, h_5^{(n)}, h_6^{(n)}\}$ are determined by 
the average value of the coefficient $d_n(k)$, i.e., by 
$d_n(B_{n+1},B_n)=a_2^{(n)}-\beta\tilde{\Phi}(B_{n+1},B_n)$ as well as the 
quantities $a_1^{(n)}$, $a_3^{(n)}$, $a_4^{(n)}$, $a_5^{(n)}$ and  
$a_6^{(n)}$. The components of the set of intermediate arguments  
$\eta^{(n)}=\{\eta_1^{(n)}, \eta_2^{(n)}, \eta_3^{(n)}, \eta_4^{(n)}, 
\eta_5^{(n)}, \eta_6^{(n)}\}$ are functions of the quantities 
(\ref{smt2d15}). They are given by the formulas
\bea
& & \eta_1^{(n)}=-(4!s^{3d})^{1/4}C_1(h^{(n)})C_4^{-1/4}(h^{(n)}), \non
& & \eta_2^{(n)}=(6s^d)^{1/2}C_2(h^{(n)})C_4^{-1/2}(h^{(n)}), \non
& & \eta_3^{(n)}=\lp\frac{32}{3}s^d\rp^{1/4}C_3(h^{(n)}) 
C_4^{-3/4}(h^{(n)}), \non  
& & \eta_4^{(n)}\equiv 1, \non
& & \eta_5^{(n)}=\frac{(4!s^{-d})^{1/4}}{5}C_5(h^{(n)})
C_4^{-5/4}(h^{(n)}), \non 
& & \eta_6^{(n)}=\frac{(6s^{-d})^{1/2}}{15}C_6(h^{(n)})C_4^{-3/2}(h^{(n)}). 
\label{smt2d16}  
\eea
The ~special ~functions ~$C_l(h^{(n)})$ ~can ~be ~expressed ~in ~terms ~of~ 
$F_l(h^{(n)})=I_l(h^{(n)})/I_0(h^{(n)})$:
\bea
C_1(h^{(n)})&=&F_1(h^{(n)}), \non 
C_2(h^{(n)})&=&F_2(h^{(n)})-F_1^2(h^{(n)}), \non  
C_3(h^{(n)})&=&F_3(h^{(n)})-3F_2(h^{(n)})F_1(h^{(n)})+ 
2F_1^3(h^{(n)}), \non
C_4(h^{(n)})&=&-F_4(h^{(n)})+4F_3(h^{(n)})F_1(h^{(n)})+ 
3F_2^2(h^{(n)})- \non
&&-12F_2(h^{(n)})F_1^2(h^{(n)})+6F_1^4(h^{(n)}), \non
C_5(h^{(n)})&=&-F_5(h^{(n)})+5F_4(h^{(n)})F_1(h^{(n)})+ 
10F_3(h^{(n)})F_2(h^{(n)})- \non 
&&-20F_3(h^{(n)})F_1^2(h^{(n)})-30F_2^2(h^{(n)})F_1(h^{(n)})+ \non 
&&+60F_2(h^{(n)})F_1^3(h^{(n)})-24F_1^5(h^{(n)}), \non
C_6(h^{(n)})&=&F_6(h^{(n)})-6F_5(h^{(n)})F_1(h^{(n)})-
15F_4(h^{(n)})F_2(h^{(n)})+ \non
&&+30F_4(h^{(n)})F_1^2(h^{(n)})-10F_3^2(h^{(n)})+ \non
&&+120F_3(h^{(n)})F_2(h^{(n)})F_1(h^{(n)})- 
120F_3(h^{(n)})F_1^3(h^{(n)})+ \non 
&&+30F_2^3(h^{(n)})-270F_2^2(h^{(n)})F_1^2(h^{(n)})+ \non
&&+360F_2(h^{(n)})F_1^4(h^{(n)})-120F_1^6(h^{(n)}).
\label{smt2d17}
\eea
Here
\be
I_l(h^{(n)})=\ili_{-\infty}^{\infty}t^lf^{(n)}(t)~dt, \qquad 
l=0,1,2,3,4,5,6, 
\label{smt2d18}
\ee
and
\be
f^{(n)}(t)=\exp(-h_1^{(n)}t-h_2^{(n)}t^2-h_3^{(n)}t^3-h_4^{(n)}t^4-
h_5^{(n)}t^5-h_6^{(n)}t^6).  
\label{smt2d19}
\ee
The effective sextic measure density of the $(n+1)$th block structure \\ 
$\cW_6^{(n+1)}(\rho)$ is written as
\bea
\cW_6^{(n+1)}(\rho)&=&\exp\Biggl[-N_{n+1}^{1/2}a_1^{(n+1)}\rho_0- 
\half~\sli_{k\leq B_{n+1}}d_{n+1}(k)\rhok\rhomk- \non
&&-~\sli_{l=3}^6\frac{a_l^{(n+1)}}{l!N_{n+1}^{l/2-1}}~\sli_{k_1,
\ldots,k_l\leq B_{n+1}}\rho_{\ve{k}_1}\cdots\rho_{\ve{k}_l}\times \non
&&\times\delta_{\ve{k}_1+\cdots+\ve{k}_l}\Biggr],
\label{smt2d20}
\eea
where $B_{n+1}=B's^{-(n+1)}$, $d_{n+1}(k)=a_2^{(n+1)}-\btPhik$,
$a_l^{(n+1)}$ are the renormalized values of the coefficients $a'_l$ 
after the integration over the $n+1$ layers of the phase space of the CV.

The coefficients of the sextic measure densities of the $(n+1)$th and $n$th 
block structures are connected through the following general RR:
\be
u_l^{(n+1)}=s^{l-\frac{l-2}{2}d}\lbr 
-q\delta_{l-2}+(u_4^{(n)})^{l/4}Y_l(h^{(n)})\rbr, \qquad 
l=1,2,3,4,5,6.
\label{smt2d21}
\ee
We have introduced new designations here:
\bea
& & u_1^{(n)}=s^{n}a_1^{(n)}, \non
& & u_2^{(n)}+q=s^{2n}d_n(B_{n+1},B_n), \qquad u_2^{(n)}=s^{2n}d_n(0), \non 
& & u_m^{(n)}=s^{mn}a_m^{(n)}, \qquad m=3,4,5,6.
\label{smt2d22}
\eea
The quantity $q=\bar q\beta\tilde{\Phi}(0)$ determines the average value of 
the Fourier transform of the potential in the $n$th layer 
$\beta\tilde{\Phi}(B_{n+1},B_n)=\beta\tilde{\Phi}(0)-q/s^{2n}$ ($\bar q$ 
corresponds to the average value of $k^2$ on the interval ($1/s,1$]). The 
functions appearing in the RR (\ref{smt2d21}) can be defined by the 
expressions
\bea
& & Y_1(h^{(n)})=s^{-d/4}C_{1sc}(\eta^{(n)})C_4^{-1/4}(h^{(n)}), \non 
& & Y_k(h^{(n)})=(-1)^{k}s^{\frac{3k-4}{4}d}C_{ksc}(\eta^{(n)})
C_4^{-k/4}(h^{(n)}), \non  
& & k=2,3,4,5,6.
\label{smt2d23}
\eea
The functions $C_{lsc}(\eta^{(n)})$ have the form similar to 
$C_{lsc}(\eta')$ (\ref{smt2d10}) under condition that the set $\eta'$ and 
the functions $f_s(t)$, $f_c(t)$ (\ref{smt2d12}) should be replaced by 
$\eta^{(n)}$ and
\bea
F_s(t)=\sin(\eta_1^{(n)}t+\eta_3^{(n)}t^3+\eta_5^{(n)}t^5)
\exp(-\eta_2^{(n)}t^2-\eta_4^{(n)}t^4-\eta_6^{(n)}t^6), \non  
F_c(t)=\cos(\eta_1^{(n)}t+\eta_3^{(n)}t^3+\eta_5^{(n)}t^5)
\exp(-\eta_2^{(n)}t^2-\eta_4^{(n)}t^4-\eta_6^{(n)}t^6),   
\label{smt2d24}
\eea
respectively. It should be noted that the formula for $I_{2lc}(\eta')$ (see 
(\ref{smt2d11})) allows us to express the function $I_{0c}(\eta^{(n)})$ 
entering (\ref{smt2d14}) in the same way. The function $C_4(h^{(n)})$ is 
given in (\ref{smt2d17}). 

The obtained expressions make it possible to investigate the properties of 
the 3D one-component spin system in the vicinity of the critical point  
($T=T_c$, $h=0$). In the absence of an external field ($h=0$), we have 
$a_1^{(n)}=a_3^{(n)}=a_5^{(n)}=0$ and relations (\ref{smt2d21}) are 
indentical with the RR presented in \cite{k189e}. The behaviour of the 
system can also be studied at $T=T_c$ and $h\rightarrow 0$.

The components of the set of intermediate arguments $\eta^{(n)}$ and the 
functions $C_{lsc}(\eta^{(n)})$, $C_l(h^{(n)})$, $Y_l(h^{(n)})$ are 
approximated by power series in deviations of the basic arguments 
$h_i^{(n)}$ from their values at a fixed point $h_i^*$. Taking into account 
the linear deviations, we find the following forms for these series:
\bea 
A_l&=&A_{l0}\lbr 1+A_{l1}(h_1^*-h_1^{(n)})+A_{l2}(h_2^*-h_2^{(n)})+
A_{l3}(h_3^*-h_3^{(n)})+\rd \non
&&\ld+A_{l5}(h_5^*-h_5^{(n)})+A_{l6}(h_6^*-h_6^{(n)})\rbr, \qquad 
l=1,2,3,4,5,6. 
\label{smt2d25}
\eea
Here $A_l$ are the above-mentioned intermediate arguments or functions. The 
coefficients $A_{li}$ are ultimately functions of the basic arguments at a 
fixed point. It should be emphasized that we necessarily arrive at 
asymptotic expansions when the Gaussian density of measure is taken as the 
basis density. The use of the sextic non-Gaussian measure density permits 
to obtain the rapidly converging series (\ref{smt2d25}). Having 
(\ref{smt2d25}), we can write and study the approximate RR. On the basis of 
(\ref{smt2d13}) and RR, the free energy $F=-k_BT\ln Z$ can be calculated.   
 
\sect{Discussion and conclusions} 

The results obtained for 3D Ising model in an external field can be applied 
to the description of liquid-gas critical points of both a one-component 
fluid \cite{rev9295,y190} and a binary fluid mixture (see, for example, 
\cite{kpa100}). The CV method with a reference system (RS) is used for these 
systems. The expression for the partition function  contains the even and 
odd powers of the variable with the corresponding cumulants and is similar 
to (\ref{smt2d4}), (\ref{smt2d5}). Detailed investigation of the properties 
of RS cumulants in \cite{rev9295} makes it possible to transform the grand 
partition function into a functional form defined on the effective block 
lattice. The functional corresponds to the partition function of the Ising 
model in an external field. A new point in problem of the liquid-gas 
critical point as compared to the case of the Ising model is the dependence 
of the critical temperature and all coefficients on the density and chemical 
potential. The latter is equivalent to the insertion of a constant external 
field into the Ising model. To describe the effects
connected with the asymmetry for the surface of coexistence 
of the phases, one should compute the partition function using the sextic 
measure density \cite{rev9295,ypro192}.

In this paper, the 3D one-component spin system in a nonzero external field 
is investigated by the CV method in the asymmetric $\rho^6$ model 
approximation. An initial expression for the partition function of the 
system is constructed in the form of a functional with explicitly known 
coefficient functions (see (\ref{smt2d6})). The partition function is 
integrated over the layers of the CV phase space. The main feature is the 
integration of short-wave spin density oscillation modes, which is generally 
done without using perturbation theory. The short-wave modes are 
characterized by the presence of the RG symmetry and are described by the 
non-Gaussian measure density. The general RR (\ref{smt2d21}) for the 
coefficients of the even and the odd powers of the variable in the two 
adjacent sextic measure densities
are found. The new special functions appearing in 
the construction of the theory using the asymmetric $\rho^6$ model are 
considered. These functions entering the RR are represented in the form of a 
power series in small deviations of the basic arguments from their values at 
a fixed point. Representing the RR in the form of a nonasymptotic series 
relates to rejecting the traditional use of perturbation theory, which is 
based on the Gaussian measure density.

The expressions under investigation allow us correctly to trace the 
dependence of the results on a field within the framework of the CV method. 
As a consequence of the presence of an external field $h$ in the Hamiltonian 
(\ref{smt2d1}), we obtain odd powers of the variable in the expressions 
(\ref{smt2d5}) for $J(\rho)$ and (\ref{smt2d6}) for the 
partition function $Z$. The cumulants $\cM_n$ in (\ref{smt2d5})
are combinations of the quantities $x=\tanh h'$ and $y=1-\tanh^2h'$, where 
$h'=\beta h$ is a generalized field. The initial coefficients $a'_l$ 
(\ref{smt2d8}) of the partition function are determined by these cumulants 
and are depended by $s_0$ on the microscopic parameters of the system (the 
effective interaction radius $b$ and the lattice constant $c$). The 
cumulants $\cM_1$, $\cM_3$, $\cM_5$ as well as the coefficients $a'_1$, 
$a'_3$, $a'_5$ are odd functions of an external field. The quantities 
$\cM_2$, $\cM_4$, $\cM_6$ and $a'_0$, $a'_2$, $a'_4$, $a'_6$ are even 
functions of a field. The RR (\ref{smt2d21}) make it possible to determine 
the elements of the block Hamiltonians from the initial data. As follows 
from the obtained expressions, the cumulants $\cM_1$, $\cM_3$, $\cM_5$ and 
coefficients $a'_1$, $a'_3$, $a'_5$,~...~, $a_1^{(n)}$, $a_3^{(n)}$, 
$a_5^{(n)}$ vanish at $h=0$. Then the RR (\ref{smt2d21}) for the asymmetric 
$\rho^6$ model turn into the RR for the $\rho^6$ model with even powers 
only \cite{k189e}. In the case when $h\neq 0$, the measure density involves 
odd powers of the variable $\rhok$ in addition to the even powers. Although 
the CV method as well as Wilson's approach exploit the RG ideas, it is 
based on the use of non-Gaussian measure densities. This allows one to 
obtain a qualitatively new form of the RR between the coefficients of the 
block Hamiltonians. In the limiting case (corresponding to the Gaussian 
basis measure density; $h=0$) these RR reduce to the Wilson RR 
\cite{rev14674}. As was shown in \cite{rev9697}, while this limiting case 
does not allow one to perform the calculation of the expression for the free 
energy of the system, it provides reliable results for the critical 
exponents of thermodynamic characteristics. The solutions of the RR under 
consideration can be used for calculating the thermodynamic functions of the 
system and for deriving the equation of state using the method proposed in 
\cite{rev9897}.

\end{document}